\pdfoutput=1

\documentclass[11pt]{article}

\usepackage[final]{acl}

\usepackage{times}
\usepackage{latexsym}

\usepackage[T1]{fontenc}

\usepackage[utf8]{inputenc}

\usepackage{microtype}

\usepackage{inconsolata}

\usepackage{graphicx}
\usepackage{subcaption} 

\title{Resource-Efficient Reference-Free Evaluation of Audio Captions}

\author{
  Rehana Mahfuz \\
  Qualcomm Technologies, Inc.  \\
  \texttt{rmahfuz@qti.qualcomm.com} 
  \\\And
  Yinyi Guo \\
  Qualcomm Technologies, Inc. \\
  \texttt{yinyig@qti.qualcomm.com} \\
  \\\And
  Erik Visser \\
  Qualcomm Technologies, Inc. \\
  \texttt{evisser@qti.qualcomm.com} \\}

\begin{document}
\maketitle
\begin{abstract}
To establish the trustworthiness of systems that automatically generate text captions for audio, images and video, existing reference-free metrics rely on large pretrained models which are impractical to accommodate in resource-constrained settings. To address this, we propose some metrics to elicit the model's confidence in its own generation. To assess how well these metrics replace correctness measures that leverage reference captions, we test their calibration with correctness measures. We discuss why some of these confidence metrics align better with certain correctness measures. Further, we provide insight into why temperature scaling of confidence metrics is effective. Our main contribution is a suite of well-calibrated lightweight confidence metrics for reference-free evaluation of captions in resource-constrained settings.

\end{abstract}

\section{Introduction}

Automated context awareness through sensors such as microphones and cameras is being relied on for applications as diverse as home security, military surveillance and machine condition monitoring. When such a system generates unreliable content, the stakes can be high. For example, if a surveillance system mistakenly captions a woodpecker's pecks as gunshots, that could trigger a security threat warning.



The traditional way of judging the quality of generated text is to measure its overlap or similarity with one or more reference texts. This is infeasible when the model is deployed, since reference captions are unavailable. Existing reference-free metrics to evaluate generated text depend on large pretrained models, which occupy too much storage and compute for deployment in resource-constrained settings. Hence, we investigate low-compute methods to evaluate caption quality in the absence of references. Specifically, we contribute the following:
\begin{itemize}
\item We propose reference-free evaluation metrics for audio captions that alleviate the need to store and run large pretrained models.
\item We validate these metrics by treating them as confidence metrics, and assess their calibration with widely accepted correctness measures.
\item We illustrate why temperature scaling of confidences is effective.
\item We discuss how to extend some of these metrics for image captions.
\end{itemize}


\begin{figure*}[]
  \includegraphics[width=17cm]{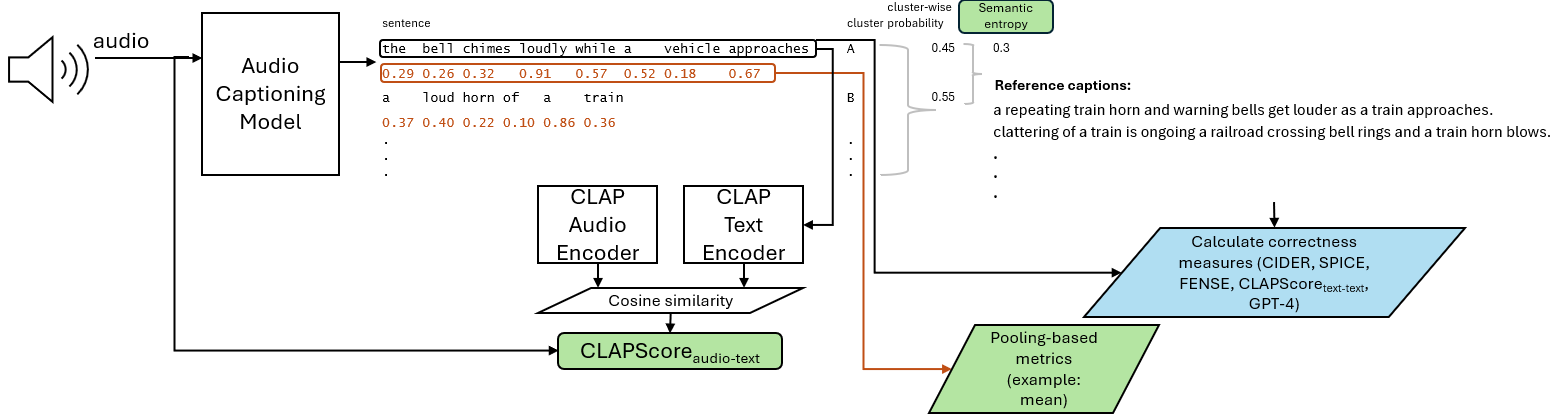}
  \caption{Our framework of obtaining confidence metrics and correctness measures.}
  \label{fig: framework_fig_a}
\end{figure*}

\begin{figure}[]
  \includegraphics[width=\columnwidth]{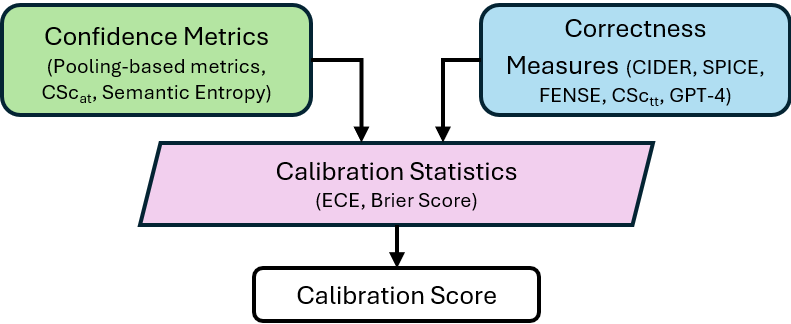}
  \caption{Our framework of measuring calibration of confidence metrics with correctness measures.}
  \label{fig: framework_fig_b}
\end{figure}

\section{Related Work}
\label{sec:related_work}

\subsection{Evaluating quality of generated text}

\subsubsection{In the presence of reference text}
\label{subsubsec: presence_of_ref_text}
There are several ways to evaluate the quality of generated text when reference text is available. BLEU \cite{bleu}, METEOR \cite{meteor} and ROUGE \cite{rouge} measure n-gram overlap, while CIDER \cite{cider} measures the cosine similarities between vectors consisting of TF-IDFs \cite{tfidf} of n-grams. SPICE \cite{spice} measures the overlap between scene graphs of reference and generated texts. BERTScore \cite{bertscore}, BLEURT \cite{bleurt2020} and FENSE \cite{fense} leverage pretrained language models in an attempt to capture semantic similarities.

\subsubsection{In the absence of reference text}
In the absence of reference text, evaluating the quality of generated text is more challenging. Often, large pretrained models are used \cite{gptscore2024, branch2024, calibrating2024, tigerscore2024, instructscore2023, t5score2023, mehri2023, g-eval2023, verbalize2023, unieval2022, bartscore2021, liu-naturalness2021, usr2020, pang2020}. The effort to transfer these evaluation capabilities to smaller models \cite{x-eval2024, minds2024} is nascent. Moreover, these pretrained models have an inherent bias to favor generations from models like themselves \cite{narcissistic2024, g-eval2023}, and can be biased against higher-quality outputs, including those written by humans \cite{limitations2022}. Further, these metrics may rely on spurious correlations with measures such as word overlap, perplexity, and length \cite{spurious2022}, may be confused by truncation errors, and errors in certain locations in the generation \cite{blind2023}. Also, verbalized confidences are not well-calibrated for difficult queries and object counting \cite{overconfidence2024}.

To evaluate the quality of generated text conditioned on other media such as images in the absence of reference text, large pretrained models are again commonly used. To detect hallucinations in image captions,\cite{aloha2024} used a Large Language Model (LLM) to extract groundable objects from the captions and measured their semantic similarities with objects detected in the image. Another method is to repeatedly generate an image from the generated caption using a large model followed by captioning the generated image to find a semantic drift indicating lack of coherence \cite{bypass-annotation2024}. These large models occupy a huge amount of space and compute, which makes it difficult to deploy them in resource-constrained settings, such as on edge devices. 
This points to the need to develop a low-compute evaluation metric for captions in the absence of reference text. We explore some options for such a metric, and, by treating them as confidence metrics, further assess their alignment with correctness measures that use reference text, in the framework of calibrating confidence metrics with correctness measures.

\subsection{Calibration} 
When a model is subjected to unseen data, a confidence metric is useful to the indicate reliability of the model's output. It is common to measure $calibration$ of the confidence metric with correctness measures that rely on the ground truth. One calibration statistic is the Expected Calibration Error (ECE) \cite{ece}, which partitions confidences into equally spaced bins, and then computes a weighted average of the absolute differences between the confidences and the correctnesses in each bin, where the weight is determined by the number of confidences in that bin. Another calibration statistic is the Brier Score \cite{brier} is the Mean Squared Error between confidence and correctness. A lower value is better for both calibration statistics.

\section{Procedure}
\label{sec: procedure}
We consider an audio captioning model $AC$ which, conditioned on an audio clip $a$, generates text $t = [t_1, t_2, ..., t_n]$, where $t_i$ is the $i^{th}$ token. 
Let $p = [p_1, p_2, ..., p_n]$ be the list of respective token probabilities. In this section, we will describe the confidence metrics we developed for audio captions. All of these metrics are deployed during inference and do not require any interference during training. An overall framework diagram is shown in Figure \ref{fig: framework_fig_a}.
\subsection{Pooling-based metrics}
\label{subsec: pooling-based-metrics}
To calculate the confidence of the generated text, we pool probabilities of the generated tokens.
We define the arithmetic mean of the token probabilities, henceforth referred to as $AM(t)$ or simply as AM as 
\begin{equation} \label{eq: am}
    AM(t) = \frac{1}{n}\sum_{i=1}^{n}p_i
\end{equation}
We define the geometric mean of the token probabilities, henceforth referred to as $GM(t)$ or simply as GM as
\begin{equation} \label{eq: gm}
    GM(t) = (\prod_{i=1}^{n}p_i)^{\frac{1}{n}}
\end{equation}
We refer to the AM and the GM as naive pooling-based confidence metrics.
We found that probabilities of non-stopword tokens carry more information, where stopwords refer to frequently occurring words which contribute little semantic value such as $a$, $and$, $is$ and $the$. Hence we also tried pooling using only probabilities of tokens which are among the $noun$, $verb$ and $adjective$ parts of speech, as judged by NLTK \cite{nltk}. Formally, let $M$ be the set of all indices identifying tokens from $t$ which are among the $noun$, $verb$ or $adjective$ parts of the speech.
We define the selective arithmetic mean of the token probabilities, henceforth referred to as $SAM(t)$ or simply as SAM as 
\begin{equation} \label{eq: sam}
    SAM(t) = \frac{1}{|M|}\sum_{i \in M}p_i
\end{equation}
We define the selective geometric mean of the token probabilities, henceforth referred to as $SGM(t)$ or simply as SGM as
\begin{equation} \label{eq: sgm}
    SGM(t) = (\prod_{i \in M}p_i)^{\frac{1}{|M|}}
\end{equation}
We refer to the SAM and the SGM as selective pooling-based confidence metrics. The collection of naive pooling-based confidence metrics and selective pooling-based confidence metrics is referred to as pooling-based metrics.

\subsubsection{Temperature Scaling}
We optionally apply temperature scaling to the list of logits $q = [q_1, q_2, ...., q_n]$ corresponding to token probabilities before the softmax layer. To clarify, the relationship between $p$ and $q$ is $p=softmax(q)$. Using a scalar temperature $temp$, the temperature-scaled probabilities $p'(temp) = [p'_1(temp), p'_2(temp), ..., p'_n(temp)]$ are obtained from $q$ as follows:

\begin{equation}
    \label{eq: temp-scaling}
    p'_i(temp) = \frac{e^{\frac{q_i}{temp}}}{\sum_{q_j \in q}e^{\frac{q_j}{temp}}}
\end{equation}

\subsection{CLAPScore}
\label{subsec: clapscore}

To measure how similar the generated text $t$ is to the audio $a$, we measure the cosine similarity between them in the multimodal space enabled by the CLAP \cite{clap} training mechanism. The $CLAPScore_{at}$ or CSc$_{at}$ is defined as
\begin{equation} \label{eq: clapscore_at}
    CSc_{at}(a,t) = \frac{ad\_emb(a).tx\_emb(t)}{||ad\_emb(a)||||tx\_emb(t)||}, 
\end{equation}
where $ad\_emb$ and $tx\_emb$ are both unary functions that project their audio and text inputs respectively into a shared multimodal space.


\subsection{Semantic Entropy}
\label{subsec: sem-entr}
To measure how consistent the model's responses are across generations for the same input, we adapted the concept of semantic entropy \cite{semantic-entropy} for audio captions. For an audio clip $a$, we sample a set of $p$ generations $T = {t^{(1)}, t^{(2)}, ..., t^{(p)}}$.
For two text generations $q$ and $r$, let us define the $CLAPScore_{tt}$, abbreviated as $CSc_{tt}$ as 
\begin{equation}
\label{eq: clapscore_tt}
    CSc_{tt}(q,r) = \frac{tx\_emb(q).tx\_emb(r)}{||tx\_emb(q)||||tx\_emb(r)||}
\end{equation}
Using $CSc_{tt}$ as the distance metric, we form clusters within T such that $t^{(u)}$ and $t^{(v)}$ belong to the same cluster if and only if $CSc_{tt}(t^{(u)}, t^{(v)}) > h$, where $h \in [-1, 1]$ is a scalar threshold.
Next, for the $l^{th}$ cluster $c_l$, we calculate its probability $P(c_l)$ as the average of probabilities of all its generations, where the probability of a generation is simply the sum of all token probabilities.

\begin{equation}
    \label{eq: cluster-prob}
    P(c_l) = \frac{1}{|c_l|}\sum_{t^{(j)} \in c_l} \sum_{i=1}^{|t^{(j)}|} t_i^{(j)}, 
\end{equation}
where $|c_l|$ is the number of generations in $c_l$, and $|t^{(j)}|$ is the number of tokens in $t^{(j)}$.
Finally, we calculate the semantic entropy $SE(T)$, also referred to as SE as 

\begin{equation}
    \label{eq: sem-entr}
    SE(T) = -\sum_{c_i \in T}P(c_i)log(P(c_i))
\end{equation}
In our experiments, $p=7$. To stay consistent with the trend among confidence metrics of being constrained between 0 and 1, with higher being better, we define the inverse semantic entropy $ISE(T)$, also referred to as ISE as

\begin{equation}
    ISE(T) = 1 - min(SE(T), 1) 
\end{equation}

\section{Experiments}
In this section, we describe the experiments to measure how well the confidence metrics we described in Section \ref{sec: procedure} calibrate with correctness measures, also shown in Figure \ref{fig: framework_fig_b}. 

\textbf{Correctness Measures:} Apart from the traditional correctness measures CIDER and SPICE and the pretrained model-based correctness measure FENSE, all of which were introduced in Subsubsection \ref{subsubsec: presence_of_ref_text}, we use two other correctness measures to judge the correctness of the generated text with respect to a reference text. The first new correctness measure is the the $CLAPScore_{tt}$ or CSc$_{tt}$ defined in Equation \ref{eq: clapscore_tt}, which calculates the cosine similarity between the generated text and the reference text in the audio-text multimodal space. The second is GPT-4's judgment regarding how well the generated text describes the audio which is described by the reference text.
To study their relationships of correctness measures with each other, we calculated the Pearson correlations between them.

\textbf{Model architecture:} Following \cite{xinhao_paper}, our audio captioning model consists of a CNN10 PANN encoder \cite{pann} followed by four layers of a transformer decoder with two heads each, with a hidden size of 256 and a feedforward dimension of 2048.
It uses text embeddings from the bert-L12-H256 model\cite{bert-l12-h256}. 

\textbf{Datasets:} We trained this model with our own audio captioning dataset (details in Appendix \ref{appendixA}). 
For evaluation, we used the evaluation splits of the AudioCaps \cite{audiocaps} and Clotho \cite{clotho} datasets, which have 957 and 1045 samples respectively. 
To find the optimum temperature for calibration, we used the validation splits of these datasets, which have 495 and 1045 samples respectively.

\textbf{Measuring Calibration:} The ECE and the Brier Score are used to measure the calibration of our confidence metrics with correctness measures. 
For pooling-based metrics, we also test the effectiveness of temperature scaling in improving calibration by selecting from $temp \in {0.1, 0.2, ..., 2.0}$ using the validation split.

\begin{table*}[]
    \centering
    \begin{tabular}{|p{2.1cm}|p{1cm}|p{0.9cm}|p{1cm}|p{1cm}|p{1.1cm}||p{1cm}|p{0.9cm}|p{1cm}|p{1cm}|p{1.1cm}|}
    \hline
    
        & \multicolumn{5}{|c|}{\textbf{Brier Score ($\downarrow$)}} & \multicolumn{5}{|c|}{\textbf{Expected Calibration Error ($\downarrow$)} }\\ \hline
        \multicolumn{11}{|c|}{\textbf{AudioCaps}} \\ \hline
        &\textbf{CIDER}&\textbf{SPICE}&\textbf{FENSE}&\textbf{CSc$_{tt}$}&\textbf{GPT-4}&\textbf{CIDER}&\textbf{SPICE}&\textbf{FENSE}&\textbf{CSc$_{tt}$}&\textbf{GPT-4} \\ \hline
        \textbf{AM}&0.24&0.2&\textbf{0.04}&0.11&\textbf{0.08}&0.21&0.42&0.09&0.31&\textbf{0.08} \\ \hline
        \textbf{SAM}&0.22&0.15&0.05&0.16&0.09&0.16&0.36&0.11&0.37&0.11 \\ \hline
        \textbf{GM}&0.22&0.16&\textbf{0.04}&0.14&0.08&0.16&0.37&\textbf{0.08}&0.36&0.09 \\ \hline
        \textbf{SGM}&\textbf{0.2}&\textbf{0.12}&0.05&0.19&0.1&\textbf{0.11}&\textbf{0.32}&0.12&0.41&0.14 \\ \hline
        \textbf{CSc$_{at}$}&0.3&0.31&0.06&\textbf{0.04}&\textbf{0.08}&0.34&0.55&0.15&\textbf{0.18}&0.12 \\ \hline
        \textbf{ISE}&0.6&0.71&0.25&0.05&0.25&0.62&0.82&0.44&\textbf{0.18}&0.41 \\ \hline

        \multicolumn{11}{|c|}{\textbf{Clotho}} \\ \hline
        \textbf{AM}&0.2&0.21&0.05&0.1&0.08&0.35&0.45&0.12&0.29&0.11 \\ \hline
        \textbf{SAM}&0.15&0.16&0.05&0.15&0.08&0.28&0.38&0.11&0.35&0.1 \\ \hline
        \textbf{GM}&0.16&0.16&\textbf{0.04}&0.14&\textbf{0.07}&0.28&0.38&\textbf{0.09}&0.35&\textbf{0.07} \\ \hline
        \textbf{SGM}&\textbf{0.12}&\textbf{0.12}&0.05&0.19&0.08&\textbf{0.22}&\textbf{0.32}&0.11&0.41&0.11 \\ \hline
        \textbf{CSc$_{at}$}&0.31&0.36&0.1&\textbf{0.03}&0.12&0.49&0.59&0.25&\textbf{0.14}&0.23 \\ \hline
        \textbf{ISE}&0.65&0.74&0.32&0.06&0.33&0.73&0.83&0.5&0.21&0.49 \\ \hline
        
    \end{tabular}
   \caption{Calibration scores on the evaluation splits of AudioCaps and Clotho with no temperature scaling.}
    \label{table:calibration_results}
\
\end{table*}

\begin{table}[]
    \centering
    \begin{tabular}{|p{0.65cm}|p{0.5cm}|p{0.45cm}|p{0.55cm}|p{0.6cm}|p{0.55cm}||p{0.5cm}|p{0.5cm}|}
    \hline
    

        &\textbf{CID ER}&\textbf{SPI CE}&\textbf{FEN SE}&\textbf{CSc$_{tt}$}&\textbf{GPT-4}&\textbf{Avg w/o TS}&\textbf{Avg w/ TS} \\ \hline
        \multicolumn{8}{|c|}{\textbf{AudioCaps}} \\ \hline
        \textbf{AM}&.2&.014&\textbf{.041}&\textbf{.008}&\textbf{.075}&.132&.068  \\ \hline
        \textbf{SAM}&.202&\textbf{.012}&.048&.013&.083&.131&.072  \\ \hline
        \textbf{GM}&.193&.016&\textbf{.041}&.009&\textbf{.075}&.127&.067  \\ \hline
        \textbf{SGM}&\textbf{.191}&.013&.049&.014&.084&.131&.07  \\ \hline

        \multicolumn{8}{|c|}{\textbf{Clotho}} \\ \hline
        \textbf{AM}&.08&\textbf{.007}&.044&\textbf{.01}&\textbf{.069}&.128&.042 \\ \hline
        \textbf{SAM}&.079&\textbf{.007}&.049&.017&.077&.118&.046 \\ \hline
        \textbf{GM}&.076&\textbf{.007}&\textbf{.042}&.012&\textbf{.069}&.113&.041 \\ \hline
        \textbf{SGM}&\textbf{.073}&\textbf{.007}&.05&.02&.076&.11&.045 \\ \hline

    \end{tabular}
   \caption{Brier scores ($\downarrow$) on the evaluation splits of AudioCaps and Clotho when using Temperature Scaling (TS).}
    \label{table: calib-w-temp}
\
\end{table}
\vspace{-2mm}

\begin{figure}[]
     \centering
     \begin{subfigure}[b]{0.23\textwidth}
         \centering
         \includegraphics[width=\textwidth]{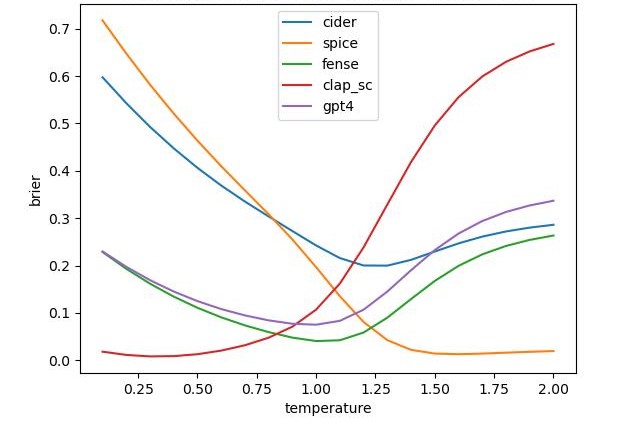}
         \caption{Arithmetic Mean}
         \label{sfig:mean-calib}
     \end{subfigure}
     \begin{subfigure}[b]{0.23\textwidth}
         \centering
         \includegraphics[width=\textwidth]{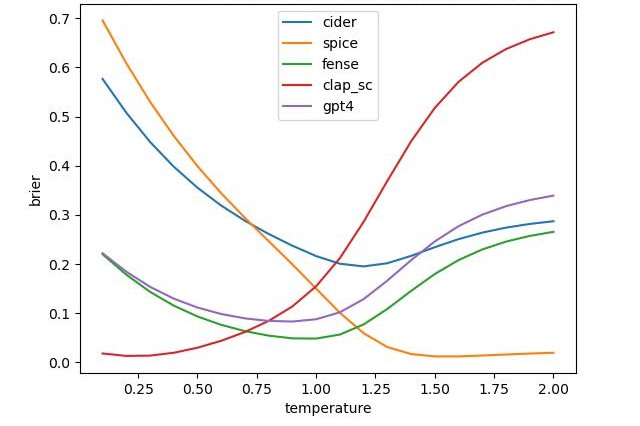}
         \caption{Selective Arithmetic Mean}
         \label{sfig:sel-mean-calib}
     \end{subfigure}
     \begin{subfigure}[b]{0.23\textwidth}
         \centering
         \includegraphics[width=\textwidth]{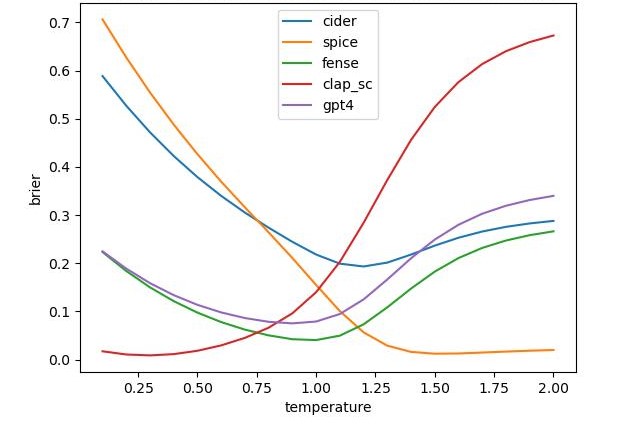}
         \caption{Geometric Mean}
         \label{sfig:geo-mean-calib}
     \end{subfigure}
     \begin{subfigure}[b]{0.23\textwidth}
         \centering
         \includegraphics[width=\textwidth]{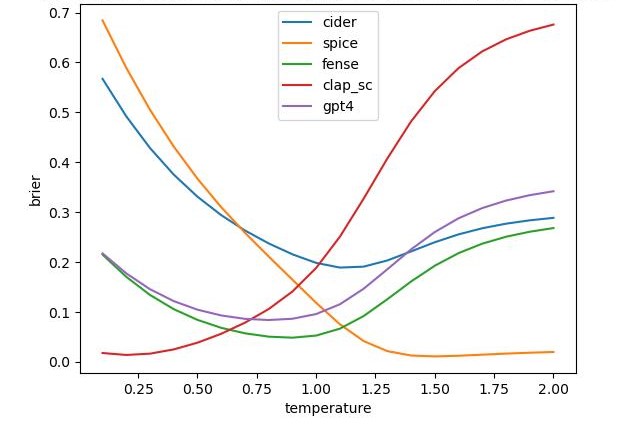}
         \caption{Selective Geometric Mean}
         \label{sfig:sel-geo-mean-conf-calib}
     \end{subfigure}

        \caption{Brier scores over temperatures for the AudioCaps dataset. Each plot shows the variation of all correctness measures over temperatures for a single confidence metric.}
        \label{fig: calib_over_temp}
\end{figure}

\begin{figure}[]
  \includegraphics[width=\columnwidth]{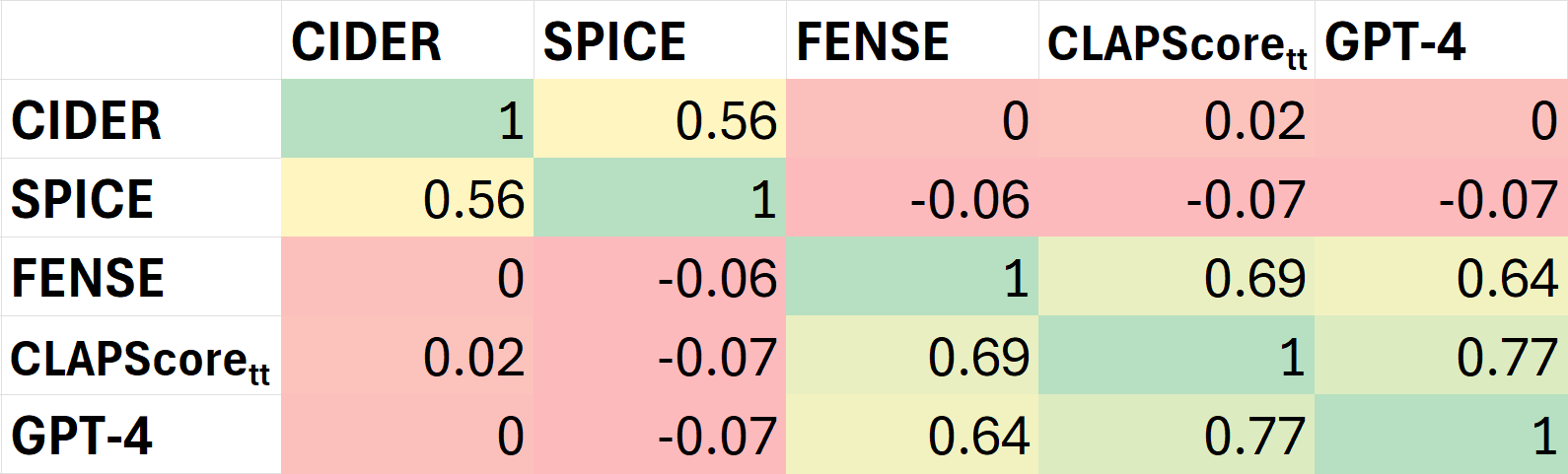}
  \caption{Pearson correlation between correctness measures for the AudioCaps dataset.}
  \label{fig :pearson_corr_bw_correctnesses}
\end{figure}

\section{Results}
\label{sec:results}

     

\begin{figure}[]
  \includegraphics[width=\columnwidth]{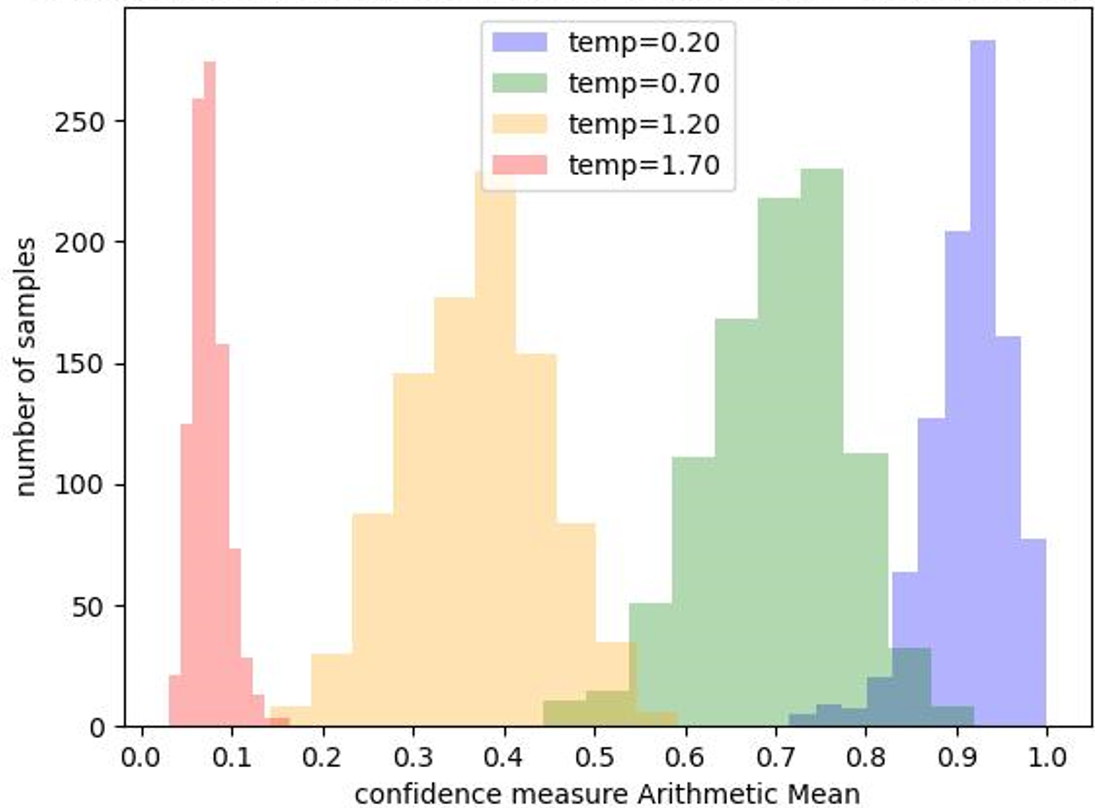}
  \caption{Variation of distribution over temperatures of the Arithmetic Mean confidence metric for the AudioCaps dataset.}
  \label{fig: vary-mean-w-temp2}
\end{figure}



\begin{figure}[]
  \includegraphics[width=\columnwidth]{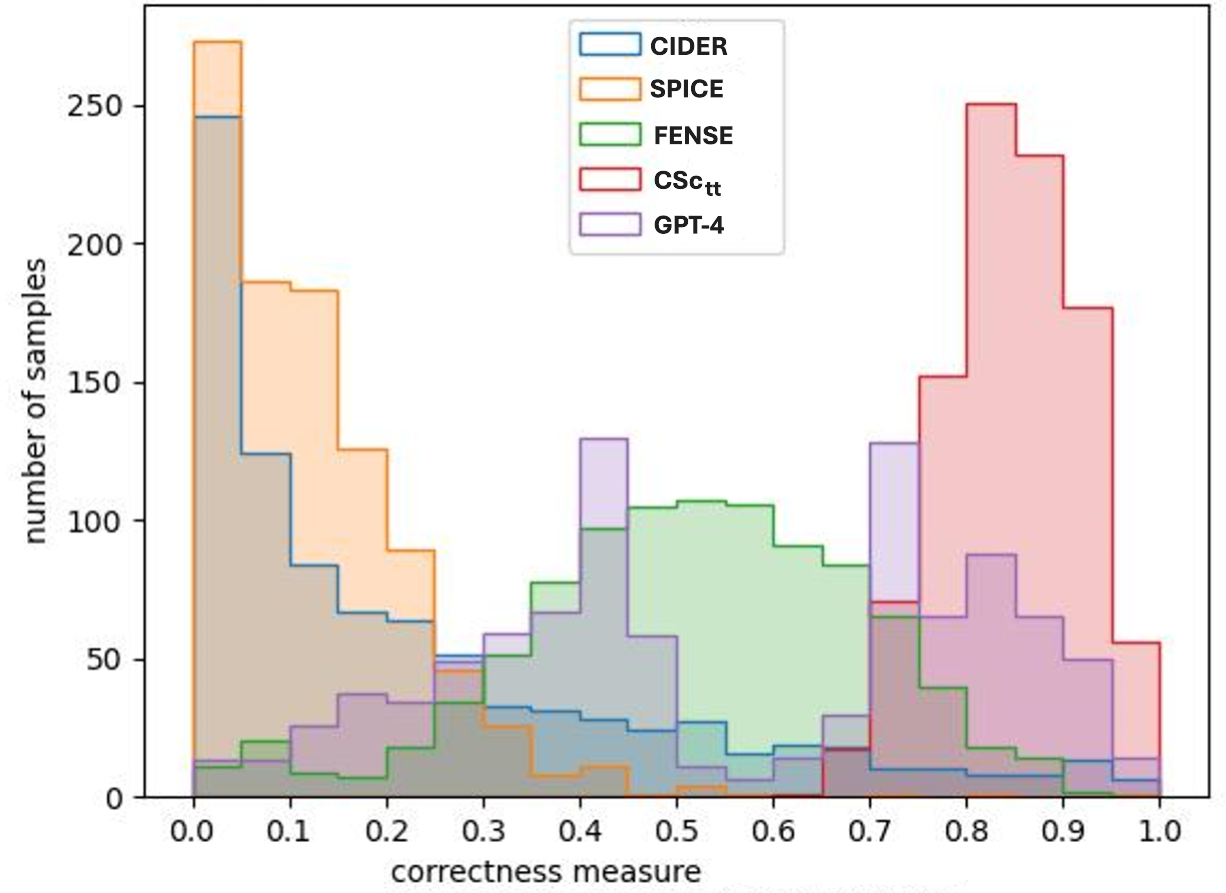}
  \caption{Distributions of correctness measures for the AudioCaps dataset.}
  \label{fig: corr-hists2}
\end{figure}

\subsection{Identifying clusters in correctness measures}
Pearson correlations among correctness measures are shown in Figure \ref{fig :pearson_corr_bw_correctnesses} for the evaluation split of the AudioCaps dataset. We observe that the traditional correctness measures CIDER and SPICE correlate with each other, while the model-based correctness measures FENSE, CLAPScore$_{tt}$ and GPT-4 correlate with each other. The same trend is observed for the Clotho dataset. 

\subsection{Evaluating Confidence Metrics} \label{sec: eval-conf-meas}

Table \ref{table:calibration_results} shows calibration scores for both datasets using the Brier Score and ECE, when no temperature scaling is used. We observe that pooling-based confidence metrics align well with all correctness measures. For traditional correctness measures CIDER and SPICE, selective pooling achieves a clear improvement over conventional pooling. This may be because CIDER gives less weight to more frequently occurring tokens, which are also the ones ignored by selective pooling, and the SPICE metric also ignores stopwords, except when needed to determine relations.
For model-based correctness measures FENSE, CLAPScore$_{tt}$ and GPT-4, pooling-based confidences continue to perform well. However, using selective pooling does not have an advantage here because model-based metrics look at the entire sentence, part of which we are discarding when using selective pooling.

For the CSc$_{tt}$ correctness measure, we can achieve even better calibration using the CSc$_{at}$ and ISE confidence metrics. 
This may be because just like the CSc$_{tt}$ correctness measure is lenient in allowing acoustically similar but semantically distant cross-triggers (examples: machine whirring and helicopter flying, typing and object clattering), the CSc$_{at}$ also forgives such cross-triggers because the cosine similarity is measured in the audio-text multimodal space.
The ISE also overlooks such cross-triggers because the CSc$_{tt}$ is used to judge consistency between the model's responses, while forming clusters to calculate the entropy.

\subsection{Effect of Temperature Scaling}

The curves of calibration quality over temperatures are shown in Figure \ref{fig: calib_over_temp} for the validation split of the AudioCaps dataset. 
The optimal temperature to calibrate a particular confidence metric with a correctness measure stays the same for both datasets, indicating its generalizability to unseen data.

Table \ref{table: calib-w-temp} shows calibration results at these optimal temperatures, on the evaluation splits of both datasets. The effectiveness of applying temperature scaling is quite pronounced, as evident from the last two columns of the table which show the calibration scores averaged over correctness measures before and after temperature scaling respectively. The average Brier score for each confidence metric almost halves after using temperature scaling. It is however important to remember that for such a dramatic improvement in calibration to be achieved, a validation set is needed to carefully select the optimal temperature. In cases where such a validation set is not available, selective pooling-based confidence metrics are still the best choice to calibrate well with traditional correctness measures.
No one pooling-based metric stands out in performance if temperature scaling is applied, an explanation for which is provided in the next Subsubsection.

\subsubsection{Why does temperature scaling work?}
\label{subsubsec: why-temp-scaling-works}

Figure \ref{fig: corr-hists2} shows the distributions of correctness measures, while Figure \ref{fig: vary-mean-w-temp2} shows how the distribution of a representative pooling-based confidence metric AM changes over temperature. 
A low temperature causes the AM to shift to the right, which matches most closely with CSc$_{tt}$, explaining why a low temperature is needed to calibrate well with CSc$_{tt}$. Similarly, a high temperature causes the AM to shift to the left, resulting in a distribution similar to those of CIDER and SPICE, explaining why a high temperature is needed to calibrate well with theses two correctness measures. Finally, a moderate temperature allows the confidence metrics's distribution to be centered around 0.5, which matches the distributions of FENSE and GPT-4, explaining why a temperature close to 1 is reasonable for calibrating with these correctness measures.



This ability of pooling-based metrics to adjust their distributions to match with those of correctness measures somewhat compensates for the differences in their computations, resulting in all of them being comparably effective.

\section{Conclusion}
\label{sec: conclusion}

We propose some resource-efficient reference-free evaluation metrics for audio captions, and validate their effectiveness by measuring their calibration with well-established correctness measures that use references. Finally, we demonstrate the effectiveness of temperature scaling and explain why it is effective. Our work enables the reliable deployment of audio captioning systems in resource-constrained settings. 


\section{Discussion}

\textbf{Feasibility of deploying our metrics in a resource-constrained setting:} While pooling-based metrics require no additional pretrained models, the CSc$_{at}$ confidence metric uses GPT-2 (126.38M) as the text encoder and HTSAT (159.45M) as the audio encoder, which results in the need to store and forward-propagate through 159.45M parameters. The ISE needs only the GPT-4 text encoder. This is much lesser than sizes of models from comparable past work that have billions of parameters \cite{branch2024, calibrating2024, g-eval2023, verbalize2023, pang2020, instructscore2023, tigerscore2024}. 

\textbf{Generalizability of our methods to other modalities:} The applicability of our proposed reference-free evaluation metrics may extend to captioning systems of other modalities as well.
The cosine similarities between CLIP \cite{clip} embeddings of pairs of images and generated text may be a good indicator of the confidence of the generated image captions.
The ISE calculated by using cosine similarities between CLIP text embeddings as the clustering criterion may also be a valuable confidence metric for image captioning models.

\textbf{Cross-triggers:} Expecting an audio captioning model to distinguish between acoustically similar sounds may be unfair, in which case, the CSc$_{tt}$ is the appropriate correctness measure, and the CSc$_{at}$ and the ISE are the recommended confidence metrics. However, since end users of captioning models may not tolerate cross-triggers, pooling-based confidence metrics are more suitable.
To reduce cross-triggers, integrating information from other sensors like cameras and motion sensors can help enhance the system’s awareness, enabling advanced tasks like summarizing and answering questions. However, due to the trade-off between using more sensors and preserving privacy, using lesser sensors is still valuable.





\section{Limitations}
\label{sec: limitations}
\begin{itemize}
\item Given that our reference-free evaluation metrics were validated with respect to the existing evaluation metrics that leverage references, our validation is limited by the quality of the existing evaluation metrics and by the quality of the human-written captions that these evaluation metrics depend on.
\item The Expected Calibration Error and Brier Score are well-suited to measure the quality of calibration of confidences for classification tasks. Its suitability to measure calibration of natural language is yet to be evaluated independently.
\item Since the objective of the study was not to evaluate the quality of the captioning model, we performed experiments only with one model. It is possible, though unlikely, that these results may not be applicable to other model architectures for the same task.
\end{itemize}

\bibliography{refs}
\clearpage

\section{Appendix A: Experimental Details}
\label{appendixA}

\subsection {Dataset}
Our audio captioning dataset was collected using the same crowdsourcing method as \cite{audiocaps} by asking people to listen to an audio clip and to write one full English sentence describing its contents. The dataset has 80,000 audio clips of length 10 seconds, and three captions corresponding to each clip, which were written by three different people. Some examples of captions from our dataset are shown in Table \ref{table: dataset}.
\begin{table}[]
    \centering
    \begin{tabular}{|p{\columnwidth}|}
    \hline
    \textbf{Captions}\\
    \hline
    A series of beeps from multiple different alarms.\\
    A continuous sharp blares of a siren followed by a loud honks and horns.\\
    A vehicle with a siren is honking.\\
    \hline
    Some rustling and a person's grunting and shouting.\\
    Someone is coughing loudly and a person suddenly shouts.\\
    A woman blows sneezes and shouts.\\
    \hline
    Metals are continuously screeching.\\
    Screeching of an operating machine.\\
    Buzzing of an electric device.\\
    \hline
    A dog howls and barks as a wind instrument is playing.\\
    Dog weeping and barking while instrumental music is playing.\\
    Musical instrument playing and a dog barking and wailing.\\
    \hline
    A loud rumble of thunder as the rain falls down.\\
    Thunder and heavy rain.\\
    A heavy rainfall accompanied by a loud bang of the thunder.\\
    \hline
    A sound of an mechanical equipment tools.\\
    A machine buzzing deeply.\\
    Screeching of an operating machine.\\
    \hline
    A loud screaming shouting and cheering of people.\\
    People are shouting and clapping.\\
    The people are cheering at full blast.\\
    \hline
    A baby crying and continuous buzzing of an electronic device.\\
    A baby crying constantly and some crackling.\\
    A baby is incessantly crying.\\
    \hline
    A man snores loudly as water rushes. \\
    The water is running and the person is snoring. \\
    A person snores loudly and water starts to flow. \\
    \hline
    A loud honking of a train that is passing by. \\
    The honking horn of a series of railroad cars moving as a unit by a locomotive or by integral motors. \\
    Many cars are making loud horn noises. \\
    \hline
    Birds are tweeting and chirping simultaneously.\\
    Birds singing and whistling wonderfully.\\
    A bird is chirping and a whistle can be heard while an equipment is creating a humming sound.\\
    \hline
    Chime of a musical instrument.\\
    The bells are ringing simultaneously.\\
    A series of loud chimes and clanks of bells.\\
    \hline
    \end{tabular}
   \caption{Example captions from our audio captioning dataset.}
    \label{table: dataset}
\
\end{table}

\subsection{Confidence and Correctness Measures}
To calculate FENSE, we used the 'paraphrase-TinyBERT-L6-v2' model \cite{sentence-bert} which is default in the aac-metrics toolkit\cite{aac-metrics}. To calculate the CLAPScore$_{at}$, we used the '2023' configuration of the CLAP model from \cite{msclap-lib}, which uses GPT-2 \cite{gpt2} as the text encoder and HTS-AT \cite{htsat} as the audio encoder. The prompt to GPT-4 for judging correctness of a caption with respect to a reference is shown in Table \ref{table: gpt-prompt}. The example scores were calculated using cosine similarities between the 'all-MiniLM-L6-v2' SentenceBERT embeddings.


\begin{table*}[]
    \centering
    \begin{tabular}{|p{\textwidth}|}
    \hline
    You will be given five reference sentences to
    describe an audio scene, and a new sentence. Using
    that, please evaluate how well a new sentence
    describes the audio scene, and provide a score
    between 0 and 1. Please provide only the score,
    and no other text.
    Here are some examples:
    \\ \hline \\
    \textbf{\underline{Example 1:}}\\
    \underline{Reference sentences:}\\
    people are singing and laughing\\
    a person is singing in melodic music while surrounded by a passing vehicle\\
    a person is singing while a man is laughing a splashing of water the wind is blowing and vehicles are passing by\\
    people are singing while cars pass by and a man in laughing\\
    people are laughing and singing while vehicles are passing by\\
    \underline{New sentence:} a person is singing while the children are playing\\
    \underline{Score:} 0.548\\
    \\
   \textbf{\underline{Example 2:}}\\
    \underline{Reference sentences:}\\
    music is playing\\
    a musical effect is playing\\
    there is instrumental music playing\\
    someone is playing a musical instrument\\
    instrumental music is playing\\
    \underline{New sentence:} a musical instrument is playing\\
    \underline{Score:} 0.801\\
.\\.\\.\\

            Now it’s your turn. \\

    \hline
    \end{tabular}
   \caption{Prompt provided to GPT-4 to judge the correctness of a caption with respect to a reference.}
    \label{table: gpt-prompt}
\
\end{table*}

\end{document}